%% file: main.tex
\documentclass[sigplan,screen,nonacm]{acmart}

\setcopyright{none}
\renewcommand\footnotetextcopyrightpermission[1]{}

\usepackage{algorithm}
\usepackage{algorithmic}

\usepackage{graphicx}
\usepackage{textcomp}

\usepackage{hyperref}
\usepackage{booktabs}
\usepackage{tabularx}
\usepackage{array}
\usepackage{makecell}
\usepackage{siunitx}
\usepackage{xfp}
\usepackage{url}
\usepackage{enumitem}
\usepackage{subcaption}
\definecolor{rowblue}{RGB}{235,245,255}
\definecolor{spdorange}{RGB}{220,110,0}

\def\BibTeX{{\rm B\kern-.05em{\sc i\kern-.025em b}\kern-.08em
    T\kern-.1667em\lower.7ex\hbox{E}\kern-.125emX}}
    
\begin{document}

\title{CAIEC: \underline{C}o-\underline{A}ccelerating \underline{I}nference and \underline{E}ntropy \underline{C}oding for Learning-Based Scientific Data Compression}

\author{Wenjing Huang}
\affiliation{%
  \institution{University of Chinese Academy of Sciences}
  \city{Beijing}
  \country{China}}
\email{huangwenjing23@mails.ucas.ac.cn}

\author{Yuhan Chen}
\affiliation{%
  \institution{China University of Petroleum-Beijing}
  \city{Beijing}
  \country{China}}
\email{cyhannah1231@gmail.com}

\author{Bing Lu}
\affiliation{%
  \institution{Institute of Computing Technology, Chinese Academy of Sciences}
  \city{Beijing}
  \country{China}}
\email{lubing@ict.ac.cn}

\author{Shuhan Zhang}
\affiliation{%
  \institution{Institute of Computing Technology, Chinese Academy of Sciences}
  \city{Beijing}
  \country{China}}
\email{liscopye@gmail.com}

\author{Zedong Liu}
\affiliation{%
  \institution{Institute of Computing Technology, Chinese Academy of Sciences}
  \city{Beijing}
  \country{China}}
\email{captain.liu77@gmail.com}

\author{Hansheng Wang}
\affiliation{%
  \institution{University of Electronic Science and Technology of China}
  \city{Chengdu}
  \country{China}}
\email{wanghansheng@std.uestc.edu.cn}

\author{Zheng Wei}
\affiliation{%
  \institution{Institute of Computing Technology, Chinese Academy of Sciences}
  \city{Beijing}
  \country{China}}
\email{weizheng@ncic.ac.cn}

\author{Yida Gu}
\affiliation{%
  \institution{Institute of Computing Technology, Chinese Academy of Sciences}
  \city{Beijing}
  \country{China}}
\email{guyida@ict.ac.cn}

\author{Xingchen Liu}
\affiliation{%
  \institution{University of Chinese Academy of Sciences}
  \city{Beijing}
  \country{China}}
\email{liuxingchen232@mails.ucas.ac.cn}

\author{Zhuoru Zhang}
\affiliation{%
  \institution{Xi'an Jiaotong-Liverpool University}
  \city{Suzhou}
  \country{China}}
\email{Zhuoru.Zhang23@student.xjtlu.edu.cn}

\author{Hao Hu}
\affiliation{%
  \institution{Institute of High Energy Physics, Chinese Academy of Sciences}
  \city{Beijing}
  \country{China}}
\email{huhao@ihep.ac.cn}

\author{Shiyuan Fu}
\affiliation{%
  \institution{Institute of High Energy Physics, Chinese Academy of Sciences}
  \city{Beijing}
  \country{China}}
\email{fusy@ihep.ac.cn}

\author{Jinyang Liu}
\affiliation{%
  \institution{Shanghai Jiao Tong University}
  \city{Shanghai}
  \country{China}}
\email{jinyang.liu@sjtu.edu.cn}

\author{Guangming Tan}
\affiliation{%
  \institution{Institute of Computing Technology, Chinese Academy of Sciences}
  \city{Beijing}
  \country{China}}
\email{tgm@ict.ac.cn}

\author{Dingwen Tao}
\authornote{Corresponding author.}
\affiliation{%
  \institution{Institute of Computing Technology, Chinese Academy of Sciences}
  \city{Beijing}
  \country{China}}
\email{taodingwen@ict.ac.cn}

\renewcommand{\shortauthors}{Huang et al.}


\begin{abstract}
Although learning-based scientific lossy compression can achieve strong rate--distortion performance, the original GPU implementations deliver less than 1~GB/s, limiting their practical deployment. Closing this throughput gap requires accelerating the complete learned-compression pipeline. 
We present CAIEC, a framework that co-accelerates model inference and latent entropy coding, the two major stages of learned-compression pipelines.
On the inference side, CAIEC uses component-wise sensitivity profiling and quality-controlled precision recovery to derive a mixed-precision configuration under a user-specified quality budget. On the coding side, it introduces GA-ANS, a granularity-adaptive GPU ANS coder that combines parallel state decomposition with arithmetic and memory-access optimizations. CAIEC selects coding granularity according to the inference time of the quality-constrained precision configuration and pipelines both stages on separate GPU streams.
Across two learned compression backbones and six real-world datasets, CAIEC achieves average compression/decompression throughputs of 23.40/15.23~GB/s on an NVIDIA H100 and 16.90/11.43~GB/s on an RTX~5090, with up to 112.72$\times$/101.04$\times$ speedups over the corresponding original CompressAI implementations using PyTorch FP32 inference and CPU ANS on H100.
On CRA5-compressed ERA5 data, CAIEC increases compression/decompression throughput from 2.1/1.4 to 20.4/13.7~GB/s, with only minor accuracy degradation in an 8-day MPTRAC particle-trajectory simulation.
\end{abstract}

\maketitle

\input{tex/01_introduction}
\input{tex/02_preliminaries}
\input{tex/04_design}

\input{tex/05_evaluation}

\input{tex/06_application}

\input{tex/07_discuss}
\input{tex/08_conclusion}

\input{tex/09_ack}

\newpage

\bibliographystyle{ACM-Reference-Format}
\bibliography{sample-base}


\end{document}

%% file: tex/01_introduction.tex
\section{Introduction}

\textbf{\textit{Motivation.}} Scientific simulations and large-scale observational instruments generate massive datasets. Weather and climate applications alone produce hundreds of terabytes to tens of petabytes~\cite{climate_data1,climate_data2,climate_data3}, straining storage and processing systems. Traditional scientific lossy compressors~\cite{cuzfp,cusz,pfpl,cuszhi} use fixed predictors~\cite{sz17,lindstrom2017fpzip} or transforms~\cite{zfp,SPERR}, limiting their ability to capture complex nonlinear and multiscale structures at high compression ratios. Learning-based compressors instead learn data-adaptive representations and can achieve better rate--distortion performance. Their success in image and video compression~\cite{balle2016end,balle2016density,dcae2025,tcm2023} has motivated growing interest in scientific data~\cite{ae-sz,SRNN-SZ,han2022coordnet,lu2021compressive,han2024kd,DCINR,10.1145/3778033,caesar}.


Table~\ref{tab:intro} illustrates the quality--throughput gap on COVID. Ball\'e2016 achieves the highest CR (275.6$\times$) at 34.12~dB, but its 0.22/0.16~GB/s compression/decompression throughput is two to three orders of magnitude below the rule-based GPU compressors. Moving learned-latent coding to the GPU with AIZ raises throughput to 0.74/1.06~GB/s, but reduces CR to 55.8$\times$. Thus, learned compression requires end-to-end acceleration without sacrificing compression efficiency.

\begin{table}[t]
\centering
\caption{Comparison of compression/decompression throughputs (CMP/DECMP THR) on the COVID dataset.}

\label{tab:intro}
\setlength{\tabcolsep}{4pt}
\renewcommand{\arraystretch}{0.95}
\footnotesize
\begin{tabular}{lccccc}
\hline
\textbf{Algorithm} & \textbf{Year} & \textbf{CR} & \textbf{PSNR} & \textbf{CMP THR} & \textbf{DECMP THR} \\
\hline
cuZFP\cite{cuzfp} & 2014 & 32.0 & 23.74 dB & 72.4 GB/s & 1246.3 GB/s \\
cuSZ\cite{cusz} & 2020 & 25.6 & 30.72 dB & 160.1 GB/s & 110.9 GB/s \\
PFPL\cite{pfpl} & 2025 & 28.9 & 30.72 dB & 430.9 GB/s & 455.2 GB/s \\
cuSZ-Hi\cite{cuszhi} & 2025 & 161.7 & 34.05 dB & 68.8 GB/s & 56.3 GB/s \\
\textbf{Ball\'e2016}\cite{balle2016end} & \textbf{2016} & \textcolor{black}{\textbf{275.6}} & \textcolor{black}{\textbf{34.12 dB}} & \textit{\textbf{0.22 GB/s}} & \textit{\textbf{0.16 GB/s}} \\
\textbf{AIZ}\cite{aiz} & \textbf{2026} & \textcolor{black}{\textbf{55.8}} & \textcolor{black}{\textbf{34.22 dB}} & \textit{\textbf{0.74 GB/s}} & \textit{\textbf{1.06 GB/s}} \\
\hline
\end{tabular}
\end{table}

\textbf{\textit{Limitations.}} Despite their rate--distortion (RD) advantage, learned compressors remain too slow for practical deployment. Their execution mainly consists of \textit{model inference}, which produces latent representations and probability information, and \textit{latent entropy coding}, which converts them into a bitstream (Fig.~\ref{fig:ai-compression}). End-to-end throughput depends on both stages, and accelerating either alone can expose the other as the new bottleneck. Compression components have different approximation sensitivities, limiting coarse-grained quantization, pruning, and distillation~\cite{rokh2023quantization,he2023pruning,yu2023distillation,Gao2025}. Meanwhile, many learned compressors retain CPU entropy coding~\cite{caesar-v,lee2025guaranteed,cra5,SRNN-SZ}; recent GPU coding~\cite{aiz, nvcomp, dietgpu} improves speed but can reduce compression ratio.

\textbf{\textit{Challenges.}} Coordinating the two stages raises two challenges. \textbf{First, NN components differ in low-precision sensitivity.} Uniform FP8 inference may improve throughput but noticeably degrade RD in sensitive components, requiring component-wise precision selection. \textbf{Second, GPU entropy coding trades compression efficiency for parallelism.} The studied backbones use Asymmetric Numeral Systems (ANS) for latent coding. Parallel ANS requires independent states or chunks: finer granularity increases parallelism but adds state overhead, whereas coarser granularity improves compression ratio but may underutilize the GPU. 


\textbf{\textit{Our Solution.}} CAIEC\footnote{We open-source the code at \url{https://github.com/opensdc-org/CAIEC}} co-accelerates inference and coding through sensitivity-guided mixed precision with selective precision restoration. Its Granularity-Adaptive ANS (GA-ANS) exposes a throughput--compression-ratio trade-off by adjusting channel granularity, which CAIEC selects based on inference time while overlapping both stages on separate GPU streams. 
A lightweight adaptation layer maps scientific tensors to the learned backbone.

We summarize the main contributions of this paper:
\begin{itemize}[leftmargin=*]

\item We propose component-wise mixed precision with quality-controlled recovery. It assigns FP16 or FP8 by each component's throughput--quality impact and restores precision to meet a user-specified quality budget.

\item We design GA-ANS, a granularity-adaptive GPU ANS coder~\cite{duda2013ans}. It combines parallel state decomposition with arithmetic and memory-access optimizations to expose a throughput--compression-ratio trade-off.

\item We co-design inference precision, GA-ANS granularity, and two-stream execution in CAIEC. Across six datasets, CAIEC reaches 23.40/15.23~GB/s compression/decompression throughput and up to 112.72$\times$ and 101.04$\times$ speedup on H100. A CRA5 case study further raises ERA5 compression/decompression throughput from 2.1/1.4 to 20.4/13.7~GB/s, with only minor accuracy degradation in an 8-day MPTRAC particle-trajectory simulation.
\end{itemize}

Section~\ref{sec:preliminaries} presents background and motivation; Section~\ref{sec:design} describes CAIEC; Sections~\ref{sec:eval} and~\ref{sec:applications} evaluate it and present a use case; and Sections~\ref{sec:discuss} and~\ref{sec:conclusion} discuss and conclude CAIEC.

%% file: tex/02_preliminaries.tex
\section{Background and Motivation}
\label{sec:preliminaries}

This section demonstrates learned scientific compression, existing inference- and coding-side acceleration, and the need for end-to-end co-acceleration.

\subsection{Scientific Lossy Compression}

Traditional scientific lossy compressors use fixed predictors or transforms to decorrelate data before quantization and entropy coding. Because CAIEC targets GPU-resident end-to-end compression, we focus on GPU compressors for direct performance comparison. Prediction-based methods include cuSZ-Hi~\cite{cuszhi}, using interpolation, and the cuSZp family~\cite{cuszp,huang2024cuszp2,cuszp3}, using lightweight parallel prediction and coding. Transform-based cuZFP~\cite{cuzfp} applies blockwise transforms, while PFPL~\cite{pfpl} emphasizes parallel decorrelation and coding. These methods are efficient on GPUs, but their fixed analytic operators can be less effective for complex nonlinear structures under aggressive compression~\cite{Gao2025,li2025}.

Learning-based compressors use data-adaptive representations instead of fixed decorrelation operators, including autoencoder-based transform coding~\cite{ae-sz,galletti2026physics,SRNN-SZ}, implicit representations~\cite{DCINR,han2024kd}, and diffusion models~\cite{lee2025guaranteed,li2025}. We target pretrained transform-coding pipelines that produce latents and probability information for entropy coding.
\begin{figure}[t]
  \includegraphics[width=\linewidth]{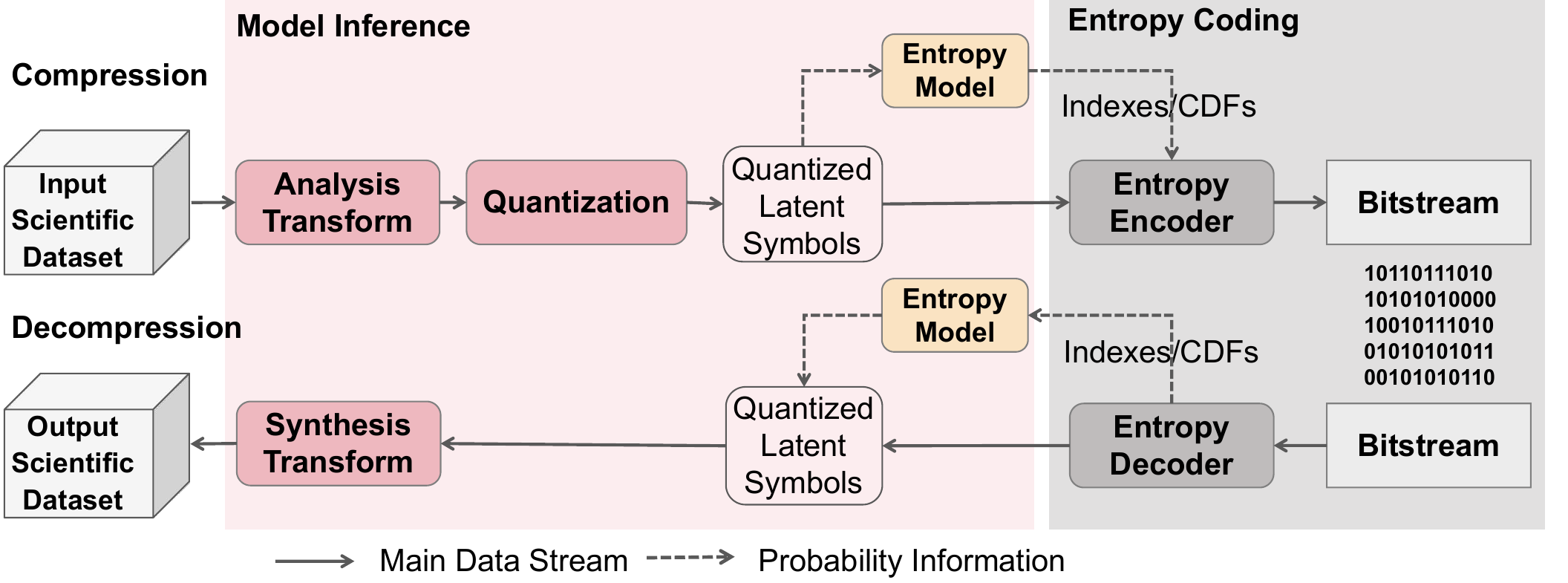}
  \caption{General pipeline of learning-based compression models.}
  \label{fig:ai-compression}
\end{figure}

Fig.~\ref{fig:ai-compression} shows CAIEC's compression pipeline. An analysis transform maps the input to quantized latent symbols; an entropy model estimates their distributions and provides indexes and CDFs; an entropy coder converts the symbols into a bitstream. Decompression reverses this process and applies a synthesis transform to reconstruct the data. We define \emph{model inference} as execution of the learned transforms and entropy-model components, and \emph{entropy coding} as bitstream generation and reconstruction from predicted distributions.

We study three models implemented using CompressAI~\cite{begaint2020compressai}, a PyTorch library for learned-compression models, entropy models, and reference entropy coding: Ball\'e2016~\cite{balle2016density}, a lightweight convolutional model with an entropy bottleneck; DCAE2025~\cite{dcae2025}, a deeper model with residual blocks, attention, and hierarchical entropy modeling; and CRA5~\cite{cra5}, a domain-specific climate compressor. Together, they span lightweight and complex generic backbones and a domain-specific deployment setting.

\begin{figure*}
  \centering
  \includegraphics[width=0.9\linewidth]{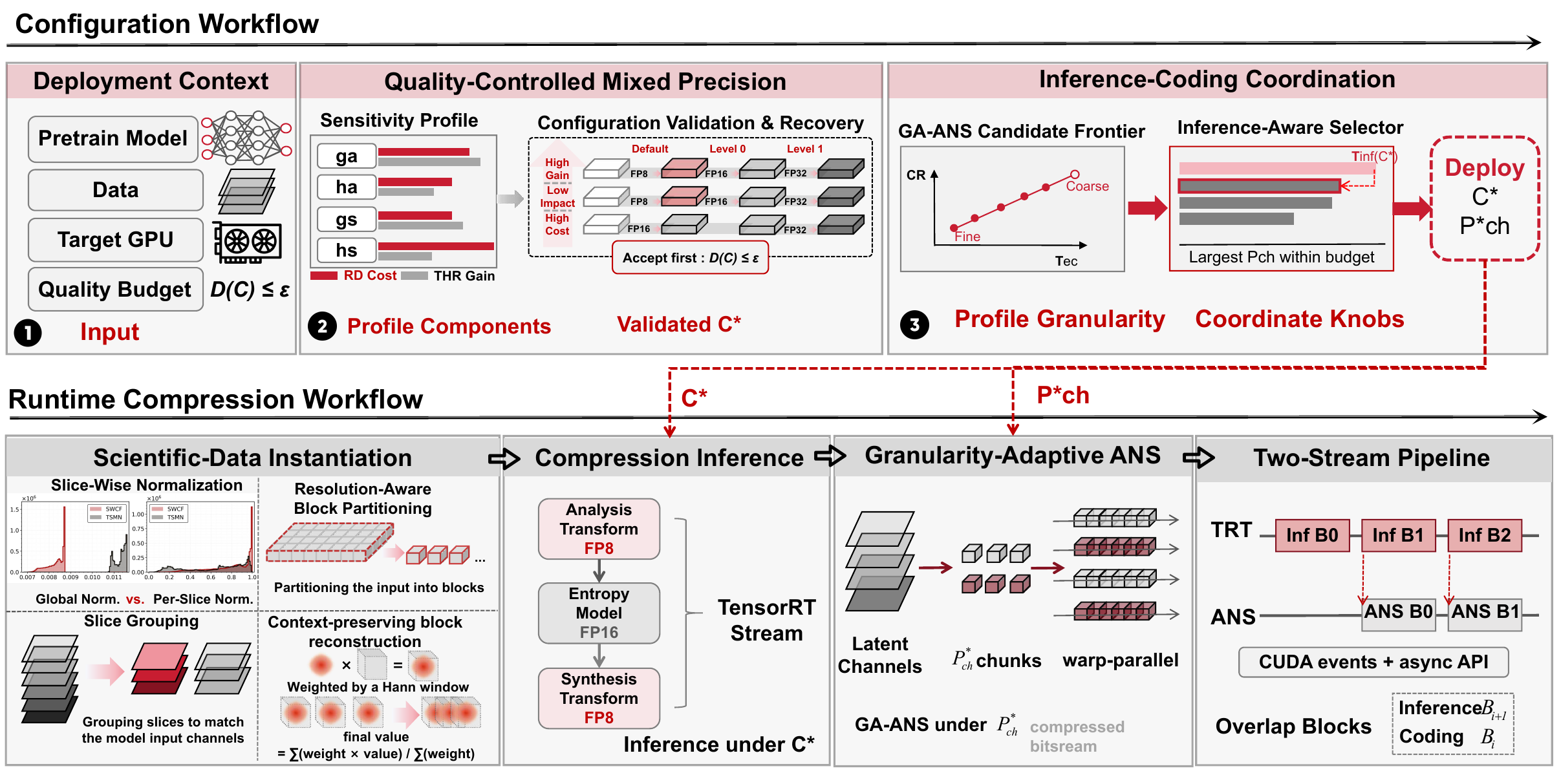}
  \caption{Overview of CAIEC. At configuration time, CAIEC derives a precision configuration $C^\star$ satisfying $D(C^\star)\leq\epsilon$, where $D(C)$ is the RD degradation of configuration $C$ and $\epsilon$ is the user-specified quality budget. It then uses the measured inference time $T_{\mathrm{inf}}(C^\star)$ to select $P_{\mathrm{ch}}^\star$, the number of latent channels per GA-ANS chunk, from the coding-time--compression-ratio frontier, denoted by $T_{\mathrm{ec}}$ and CR. At runtime, $C^\star$ and $P_{\mathrm{ch}}^\star$ control mixed-precision inference and GA-ANS, while inference of batch $B_{i+1}$ overlaps coding of batch $B_i$.}
  \label{fig:overview}
\end{figure*}

\subsection{Acceleration of Learning-Based Compression}

\noindent\textbf{Inference-side acceleration.}
Quantization, pruning, distillation, and neural architecture search reduce inference cost by lowering numerical precision, removing computation, approximating model mappings, or selecting cheaper architectures, respectively~\cite{rokh2023quantization,he2023pruning,yu2023distillation}. In learned compression, these changes can affect reconstruction quality and latent bitrate, while transform and entropy-model components have different numerical sensitivities~\cite{Gao2025}. Uniform policies therefore cannot capture each component's throughput--RD trade-off.

Learned-compression-specific methods often adapt model structure or execution paths. Slimmable Compressive Autoencoders~\cite{yang2021slimmable} and CBANet~\cite{guo2023cbanet} adjust channel width, branches, or decoding complexity to provide rate--complexity trade-offs. AdaNIC~\cite{tao2023adanic} uses input-adaptive routing to vary computation. These approaches improve flexibility, but rely on architecture- or training-specific support for execution paths.

Other work targets quantized or hardware-aware deployment. FPX-NIC~\cite{jia2022fpx} uses fixed-point quantization and FPGA implementation, while FLARE~\cite{jia2025flare} co-designs model structure, dataflow, and hardware. CAIEC is complementary to these approaches: it profiles functional components of pretrained backbones and configures their deployment precision without introducing alternative model paths.

\noindent\textbf{Coding-side acceleration.}
Another direction moves entropy coding to GPUs. Representative systems span scientific-data coders such as MANS~\cite{mans}, general-purpose GPU lossless coders such as nvCOMP~\cite{nvcomp} and DietGPU~\cite{dietgpu}, and progressive or specialized designs such as GPULZ~\cite{zhang2023gpulz} and Prism~\cite{prism}. These coders target integer, byte-stream, or progressive representations rather than the symbol-wise distributions of learned entropy models. AIZ~\cite{aiz}, the closest coding-side work to CAIEC, moves learned-latent bitstream generation to GPUs to remove the coding bottleneck.


\subsection{Limitations and Motivation}

The works above accelerate either model inference or entropy coding, but each addresses only part of the end-to-end learned-compression pipeline.

\noindent\textbf{Inference-side limitations.}
Existing inference-side techniques can reduce model execution cost, but many require retraining, architecture modifications, alternative execution paths, or specialized hardware. Moreover, learned compression differs from conventional neural inference because reduced precision affects both reconstruction quality and latent bitrate, while transform and entropy-model components can exhibit different numerical sensitivities. Existing techniques therefore do not directly provide component-wise deployment precision for an unchanged pretrained backbone under an explicit acceleration-induced RD budget.

\noindent\textbf{Coding-side limitations.}
Generic GPU coders are not designed to directly exploit the symbol-wise probabilities from learned entropy models. Learned-latent GPU coding~\cite{aiz} addresses this representation mismatch, but parallel coding still exposes a trade-off between GPU utilization and compression efficiency: finer partitioning increases parallelism but adds state/metadata overhead, whereas coarser partitioning better preserves compression ratio but may reduce coding throughput. A fixed or coding-only point cannot adapt this trade-off to backbones with different inference costs.

\noindent\textbf{Need for end-to-end co-design.}
Inference and entropy coding jointly determine end-to-end performance, and their relative costs vary across backbones and deployment configurations. Accelerating inference alone can shift the bottleneck to entropy coding, whereas accelerating coding alone can leave inference on the critical path. Independently maximizing either stage is also insufficient: aggressive precision reduction can degrade RD performance, while excessive coding parallelism can reduce compression efficiency. End-to-end acceleration therefore requires coordinating inference precision and coding granularity under a common quality objective, so that neither stage becomes the new bottleneck and compression effectiveness is preserved.


%% file: tex/04_design.tex
\section{Design of CAIEC}
\label{sec:design}

This section presents CAIEC's GA-ANS GPU entropy coding, component-wise mixed-precision inference, configuration and runtime coordination, and scientific-data instantiation.

\subsection{Overview of CAIEC}

Fig.~\ref{fig:overview} summarizes CAIEC in two phases. During configuration, CAIEC takes a pretrained model, representative data, a target GPU, and a user-specified quality budget $\epsilon$. Let $C$ denote a component-wise precision configuration and $D(C)$ its acceleration-induced RD degradation. Component-wise profiling and full-configuration recovery produce $C^\star$ satisfying $D(C^\star)\leq\epsilon$. CAIEC then measures its inference time $T_{\mathrm{inf}}(C^\star)$. On the coding side, $P_{\mathrm{ch}}$ denotes the number of latent channels per independently coded chunk, while $T_{\mathrm{ec}}(P_{\mathrm{ch}})$ and $\mathrm{CR}(P_{\mathrm{ch}})$ denote the corresponding entropy-coding time and compression ratio. From the profiled GA-ANS frontier $\{P_{\mathrm{ch}},T_{\mathrm{ec}}(P_{\mathrm{ch}}),\mathrm{CR}(P_{\mathrm{ch}})\}$, the inference-aware selector chooses $P_{\mathrm{ch}}^\star$ to preserve coding efficiency without making entropy coding the new bottleneck.

At runtime, CAIEC adapts tensors into model-compatible batches and executes inference and GA-ANS under $C^\star$ and $P_{\mathrm{ch}}^\star$, respectively. Separate TensorRT and ANS streams overlap inference of batch $i+1$ with coding of batch $i$; decompression follows the reverse workflow

\begin{algorithm}[t]
\caption{Serial CPU rANS encoding for learned latent data.}
\small
\label{alg:baseline_serial_latent_rans}
\begin{algorithmic}[1]
\REQUIRE Latent symbols $\mathbf{s}$ and entropy-model metadata $\mathcal{M}$
\ENSURE Compressed bitstream $\mathbf{bs}$
\STATE Initialize rANS state $r$
\FOR{$i = |\mathbf{s}|-1$ down to $0$}
    \STATE $(\mathit{start}, \mathit{freq}) \leftarrow \textsc{LookupInterval}(\mathbf{s}[i],\mathcal{M})$
    \STATE $r \leftarrow \lfloor r/\mathit{freq}\rfloor 2^{b_{\mathrm{cdf}}} + (r\bmod\mathit{freq}) + \mathit{start}$
\ENDFOR
\STATE Flush $r$ into $\mathbf{bs}$ and \textbf{return} $\mathbf{bs}$
\end{algorithmic}
\end{algorithm}

\subsection{GA-ANS: Granularity-Adaptive Entropy Coding}
\label{sec:gaans}
\noindent\textbf{Serial ANS and GPU challenges.}
ANS~\cite{duda2013ans} provides high compression efficiency and is widely used by learned compression backbones~\cite{caesar-v, balle2016density, dcae2025, li2025}. A conventional CPU encoder maintains one rANS state $r$ and processes symbols in reverse order because of the LIFO property of ANS. For a symbol whose quantized cumulative distribution function (CDF) interval has lower bound $\mathit{start}$, frequency $\mathit{freq}$, and precision $b_{\mathrm{cdf}}$ bits, the state update is
\[
r_{\mathrm{new}} = \left\lfloor\frac{r}{\mathit{freq}}\right\rfloor 2^{b_{\mathrm{cdf}}}
+ (r \bmod \mathit{freq}) + \mathit{start}.
\]
Algorithm~\ref{alg:baseline_serial_latent_rans} summarizes this process, using entropy-model metadata to look up each symbol's CDF interval.

\begin{figure}[t]
\includegraphics[width=0.9\linewidth]{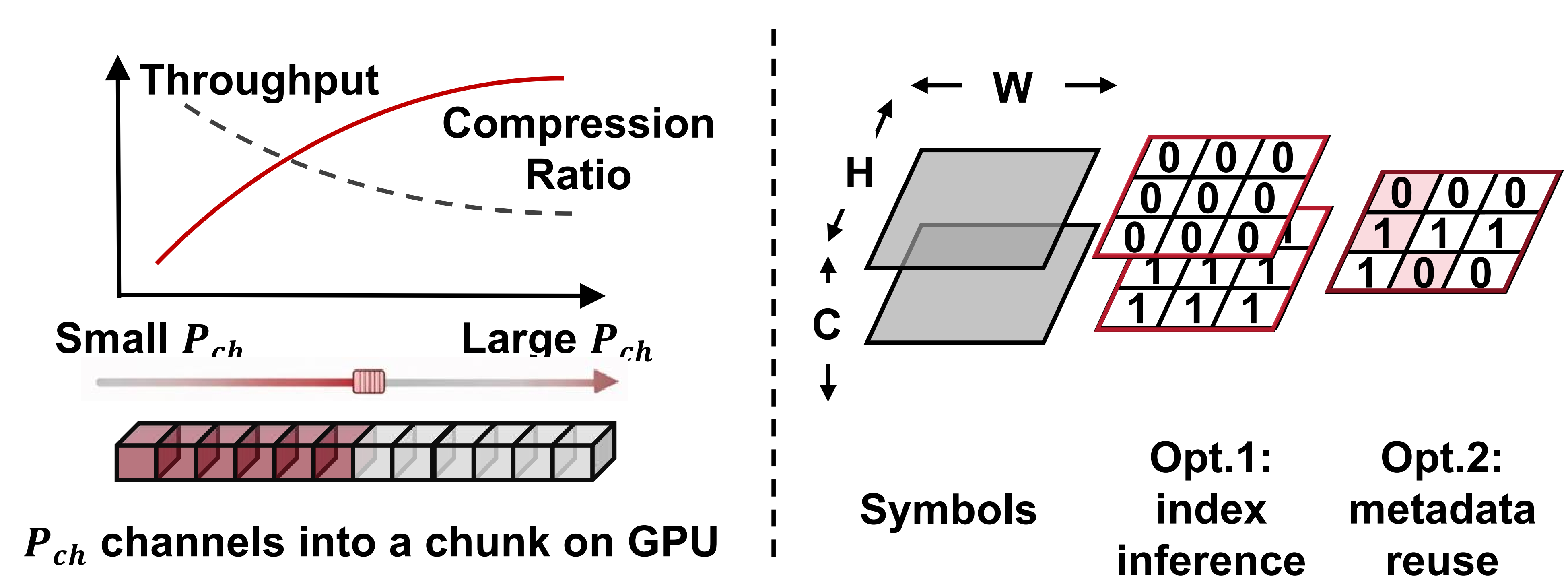}
  \caption{Throughput--CR trade-off under different $P_{\mathrm{ch}}$ settings.}
  \label{fig:P-control}
\end{figure}

This procedure is inefficient on GPUs when processing latent data. First, updates to a single state are serial, preventing direct symbol-level parallelization. Second, each symbol may require index-map and CDF-related metadata accesses. Third, the state update performs 64-bit integer division and modulo on the critical path. 
Fig.~\ref{fig:P-control} summarizes the GA-ANS optimizations: chunk- and warp-level parallelism expose coding parallelism, strength reduction removes per-symbol integer division, and metadata reuse, shared-memory CDF staging, and warp-interleaved bitstream layout improve memory access while preserving the learned probability model.

\noindent\textbf{Algorithmic optimizations.} 
\noindent\textit{Granularity-adaptive parallel coding.}
For a latent tensor, GA-ANS groups every $P_{\mathrm{ch}}$ consecutive channels into an independent chunk. The encoder writes chunk outputs to an output buffer, computes their packed offsets with a prefix sum, records the required metadata, and assembles a compact bitstream.

$P_{\mathrm{ch}}$ controls the trade-off between coding throughput and compression ratio. A smaller $P_{\mathrm{ch}}$ exposes more parallelism and improves throughput but introduces more chunk overhead, whereas a larger $P_{\mathrm{ch}}$ reduces this overhead and improves compression ratio at the cost of lower parallelism.

GA-ANS removes this coarse-granularity bottleneck by assigning one warp to each chunk. The warp lanes maintain independent rANS states and process interleaved, lane-private symbol subsequences: encoding traverses each subsequence in reverse order, while decoding traverses the corresponding positions forward. The packed chunk stores the information needed to initialize these states and restore symbols to their original positions. This intra-chunk parallelism keeps coarse chunks GPU-efficient with lower state and header overhead.

\noindent\textit{Division strength reduction.}
The rANS state update performs one 64-bit division and modulo for each encoded symbol. Because the frequency $\mathit{freq}$ is drawn from a bounded quantized CDF domain, GA-ANS precomputes exact division parameters for each valid frequency. At runtime, it replaces division with a multiply-high and shift sequence, together with a correction step when required by the precomputed scheme. Given the resulting quotient $q$, the remainder is reconstructed as $r-q\cdot\mathit{freq}$. This removes general 64-bit division from the per-symbol coding path.

\noindent\textbf{Memory-Access optimizations.}
\noindent\textit{Index inference and metadata reuse.}
Latent tensors often use channel-identity CDF indexing, allowing GA-ANS to infer each index from the symbol position and avoid per-symbol map loads. Otherwise, it caches the current index and associated table location, size, and symbol offset in registers until the index changes.

\noindent\textit{Shared-memory CDF staging.}
To accelerate the subsequent CDF-entry lookups, GA-ANS cooperatively stages the CDF tables used by a chunk in shared memory when they fit within the available capacity. Register-cached table locations then refer to the staged tables, allowing repeated accesses to CDF entries to avoid global memory.

\noindent\textit{Warp-interleaved bitstream layout.}
During encoding, each warp lane may emit multiple fixed-width words as its rANS state is renormalized. We refer to the $j$th words emitted by the lanes as output round $j$, and write $[L_i{:}W_j]$ for the $j$th word of lane $i$. Instead of storing all words from each lane consecutively, GA-ANS places words from the same round at adjacent addresses:
\[
\underbrace{[L_0{:}W_0][L_1{:}W_0]\ldots[L_{31}{:}W_0]}_{\text{round 0}}
\underbrace{[L_0{:}W_1][L_1{:}W_1]\ldots[L_{31}{:}W_1]}_{\text{round 1}}
\ldots.
\]

During decoding, lanes that require their next word in the same round therefore read adjacent locations, enabling coalesced memory accesses. Each lane tracks its own round position and consumes a word only when its state requires renormalization, so the decoder uses this layout directly without a separate de-interleaving buffer.

\subsection{Component-Wise Mixed-Precision Inference}
\label{sec:precision}

Although GA-ANS accelerates latent coding, inference remains costly for complex backbones. CAIEC therefore uses lightweight sensitivity profiling to derive a component-wise precision configuration, then validates and, if needed, recovers the full configuration under a user-defined RD budget.

\noindent\textbf{Component-wise sensitivity analysis.}
The candidate components are functional modules in the learned compressor, including the analysis transform, synthesis transform, and entropy-model or parameter-estimation modules. They play different roles in feature extraction, redundancy removal, reconstruction, and probability estimation, and consequently exhibit different numerical sensitivities. Uniformly reducing all components to FP8 can therefore cause unnecessary RD degradation. Exhaustive mixed-precision search is also impractical: for $k$ candidate components and three precision choices (FP32, FP16, and FP8), it requires evaluating up to $3^k$ complete configurations. CAIEC instead uses the all-FP16 configuration as a practical acceleration reference and probes one component at a time by reducing it from FP16 to FP8 while keeping the others at FP16. These single-component configurations are used only to estimate marginal effects, not as final deployments. Their profiles guide a joint default configuration that can assign FP8 to multiple high-gain and low-impact components; CAIEC then validates and, if necessary, restores precision in this complete configuration. Including the reference, profiling evaluates only $k+1$ configurations, reducing its complexity from exponential, $O(3^k)$, to linear, $O(k)$. We use FP32, FP16, and FP8 because FP8 has broader hardware support, while the same profiling and recovery procedure can also be applied to FP4.

\begin{figure}[t]
  \includegraphics[width=0.95\linewidth]{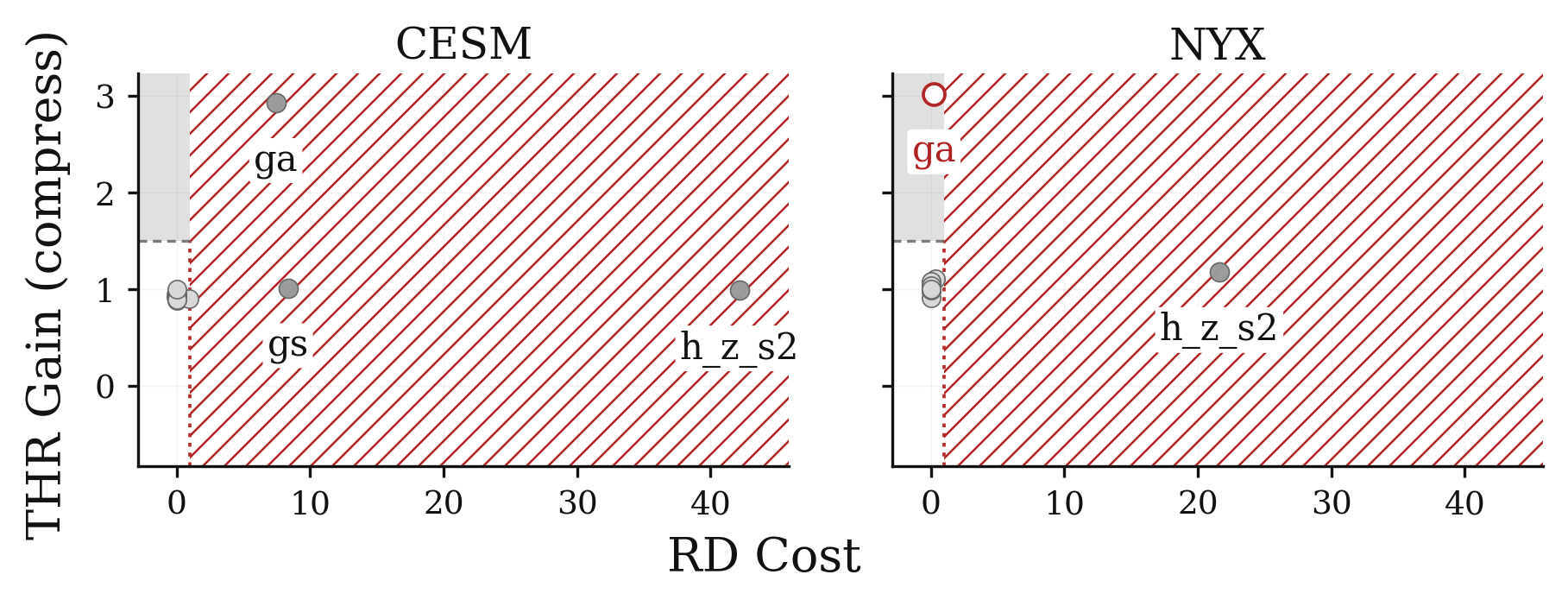}
  \caption{Component analysis results of DCAE2025 on the CESM and NYX datasets.}
  \label{fig:mix-gain-cost}
\end{figure}

\begin{figure}[t]
  \centering
\includegraphics[width=\linewidth]{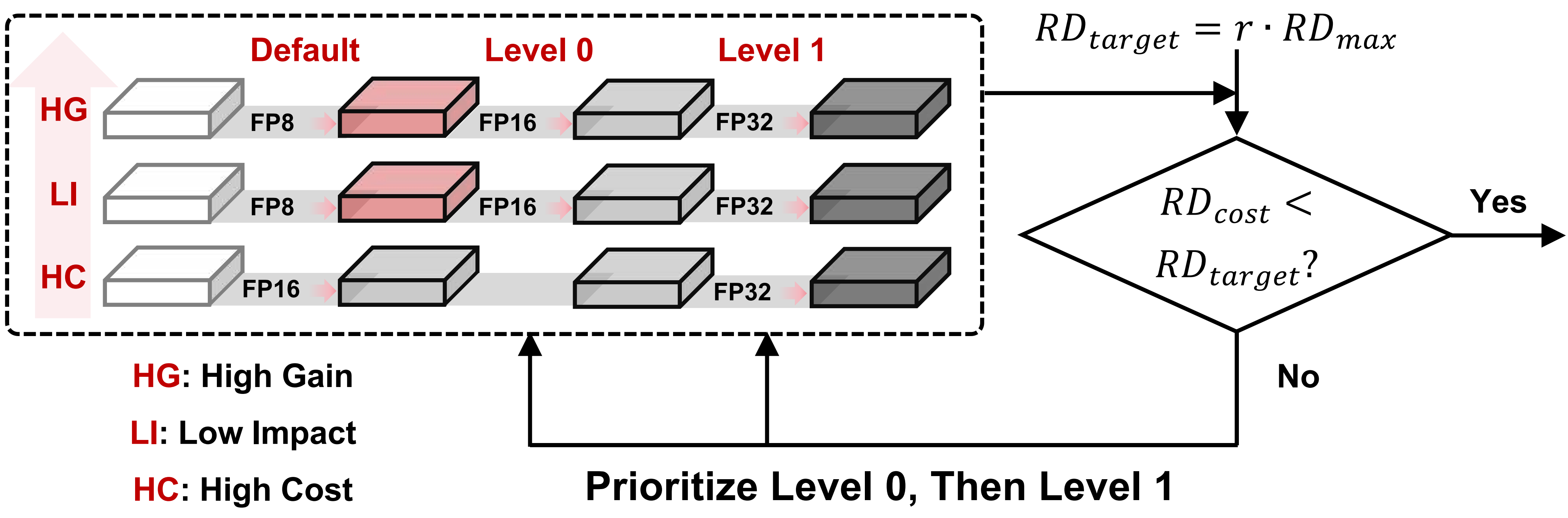}
  \caption{Quality-controlled precision recovery. CAIEC measures each complete precision configuration, progressively restoring component precision when the requested quality budget is unmet.}
  \label{fig:mp-recovery}
\end{figure}
For component $i$, CAIEC measures bitrate, reconstruction quality, and inference time of its perturbed configuration. It quantifies quality sensitivity using
\[
\mathrm{RD\ Cost}_i =
\log_{10}\frac{\mathrm{bpp}_i}{\mathrm{bpp}_{\mathrm{ref}}}
+
\left(\mathrm{PSNR}_{\mathrm{ref}}-\mathrm{PSNR}_i\right),
\]
where $\mathrm{bpp}_{\mathrm{ref}}$ and $\mathrm{PSNR}_{\mathrm{ref}}$ are the bitrate and reconstruction quality of the all-FP16 reference, and $\mathrm{bpp}_i$ and $\mathrm{PSNR}_i$ are obtained when only component $i$ uses FP8. The first term captures the relative bitrate change, while the second captures the PSNR loss. The logarithm reduces the scale disparity between the two terms. A larger $\mathrm{RD\ Cost}_i$ therefore indicates a higher risk of reducing component $i$ to FP8.

CAIEC estimates the acceleration benefit using
\[
\mathrm{THR\ Gain}_i =
1+\frac{k(T_{\mathrm{ref}}-T_i)}{T_{\mathrm{ref}}},
\]
where $T_{\mathrm{ref}}$ is the all-FP16 inference time and $T_i$ is measured when only component $i$ uses FP8. The normalized reduction $(T_{\mathrm{ref}}-T_i)/T_{\mathrm{ref}}$ is the component's marginal saving; multiplying by the number of candidate components $k$ projects this saving to a common whole-model scale, and the leading 1 sets the no-improvement baseline to one. Thus, a larger $\mathrm{THR\ Gain}_i$ indicates greater potential throughput benefit.

Both metrics are first-order profiling signals used to rank and classify components, not predictions of the RD performance or throughput of a complete mixed-precision configuration. Because interactions among simultaneously quantized components are not captured by one-at-a-time perturbation, CAIEC measures and validates every complete configuration before accepting it.

\noindent\textbf{Default mixed-precision configuration.}
CAIEC uses fixed empirical thresholds to divide the sensitivity profile into high-cost, high-gain, and low-impact groups. As illustrated in Fig.~\ref{fig:mix-gain-cost}, a component is classified as \emph{high-cost} when $\mathrm{RD\ Cost}_i>1$, indicating a noticeable RD penalty under FP8. Among the remaining components, for which $\mathrm{RD\ Cost}_i\leq 1$, those with $\mathrm{THR\ Gain}_i>1.5$ are classified as \emph{high-gain}; the rest are classified as \emph{low-impact}. We set these thresholds empirically based on recurring quality--throughput patterns observed in preliminary profiling across representative backbones and datasets, and keep them fixed when constructing the default configuration. They are conservative heuristics rather than theoretically optimal or automatically learned boundaries.

As shown in Fig.~\ref{fig:mix-gain-cost}, CAIEC assigns FP16 to high-cost components and FP8 to high-gain and low-impact components:
\[
C^{(0)}(m)=
\begin{cases}
\mathrm{FP16}, & m\in\mathcal{M}_{\mathrm{HC}},\\
\mathrm{FP8}, & m\in\mathcal{M}_{\mathrm{HG}}\cup\mathcal{M}_{\mathrm{LI}}.
\end{cases}
\]
These thresholds only construct a throughput-oriented initial configuration; they do not guarantee its final RD quality. CAIEC therefore measures the full configuration and invokes precision recovery if it violates the requested quality budget.

\noindent\textbf{Quality-controlled precision recovery.}
Starting from the default configuration, CAIEC restores low-impact and then high-gain components from FP8 to FP16, followed if necessary by high-cost components from FP16 to FP32. It processes each group by descending RD cost, remeasures the complete configuration after each change as shown in Fig.~\ref{fig:mp-recovery}.

\begin{figure}[t]
  \centering
  \includegraphics[width=0.8\linewidth]{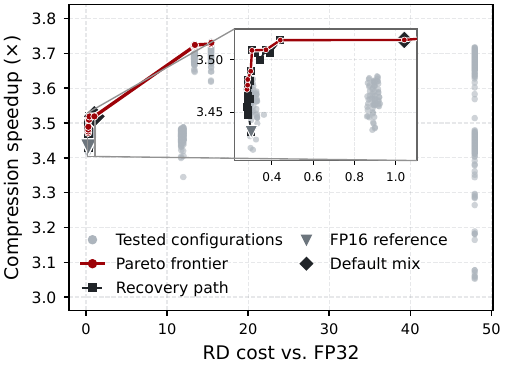}
  \caption{Precision-recovery case study on the CESM dataset.}
  \label{fig:recovery-landscape}
\end{figure}

Fig.~\ref{fig:recovery-landscape} overlays CAIEC's guided recovery trajectories on the CESM~\cite{cesm} configuration landscape. Exhaustively measuring all 2,048 configurations and their Pareto frontier takes about two days, whereas CAIEC evaluates only configurations along trajectories targeting 75\% and 50\% of the default configuration's initial RD-cost gap to all-FP32, completing in minutes. The selected configuration is then fully measured, and its inference time guides coding-granularity selection.

\subsection{Inference--Coding Co-Design}
\label{sec:codesign}

CAIEC connects mixed-precision inference and GA-ANS at two levels. At configuration time, it selects coding granularity according to the inference time of the quality-constrained precision configuration. At runtime, it pipelines inference and coding across blocks to reduce stage serialization.

\noindent\textbf{Inference-aware coding-granularity selection.}
For a selected precision configuration $C^\star$, CAIEC obtains $T_{\mathrm{inf}}(C^\star)$ from profiling. GA-ANS provides candidate coding points
\[
\left\{P_{\mathrm{ch}},\;T_{\mathrm{ec}}(P_{\mathrm{ch}}),\;\mathrm{CR}(P_{\mathrm{ch}})\right\}.
\]
Larger $P_{\mathrm{ch}}$ values generally reduce state and header overhead and therefore improve CR, but also increase coding time. CAIEC consequently chooses the largest coding granularity whose coding time fits within the inference-time budget:
\[
P_{\mathrm{ch}}^\star=
\max P_{\mathrm{ch}}
\quad\text{s.t.}\quad
T_{\mathrm{ec}}(P_{\mathrm{ch}})\leq T_{\mathrm{inf}}(C^\star).
\]

This selection makes the coupling explicit: a faster inference configuration may require finer-grained coding to prevent coding from becoming the bottleneck, whereas a slower configuration can use coarser coding and trade otherwise idle coding time for a better CR. CAIEC therefore does not independently maximize inference and coding throughput.

\begin{figure}[t]
  \centering
  \includegraphics[width=0.95\linewidth]{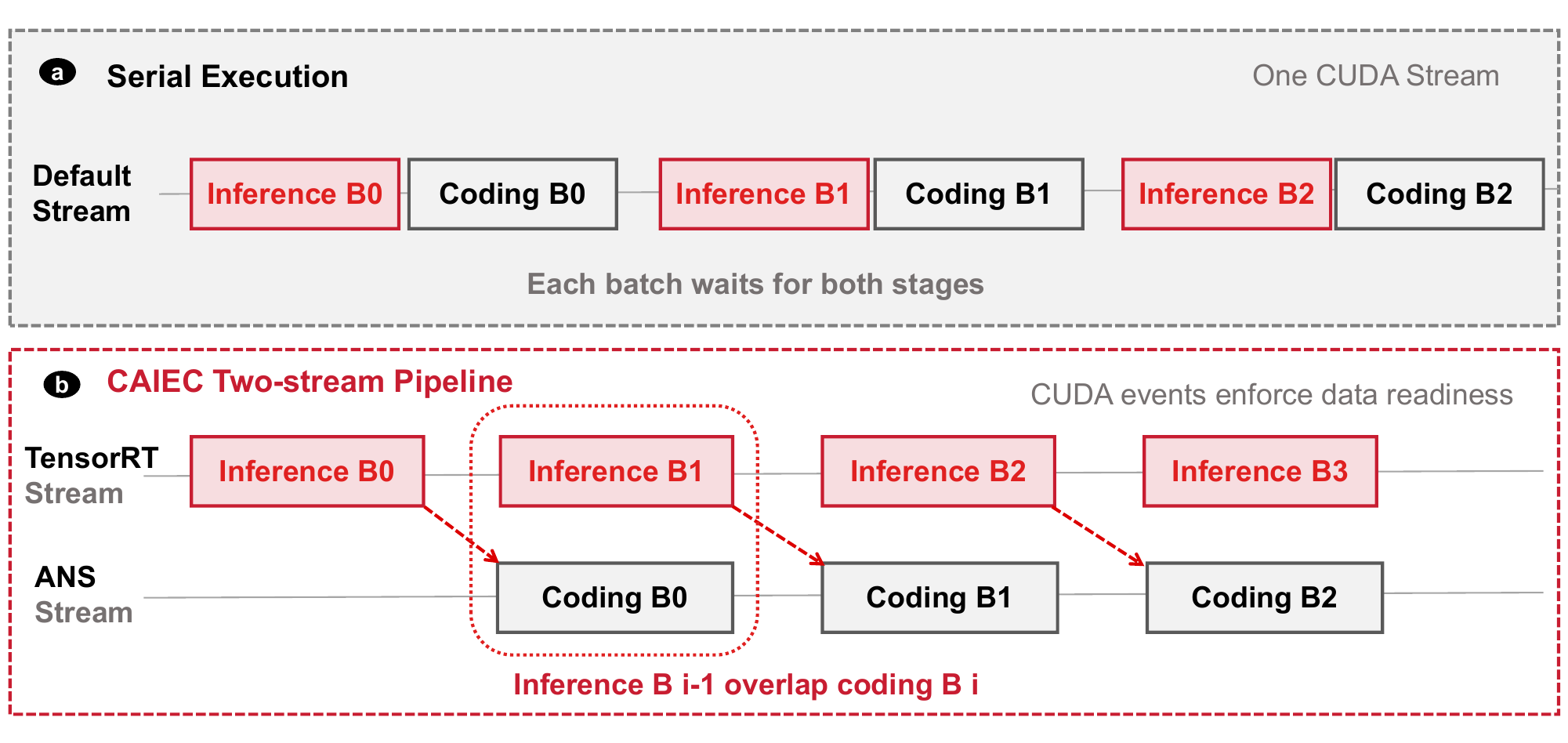}
  \caption{CAIEC two-stream inference--coding schedules.}
  \label{fig:pipeline}
\end{figure}

\begin{figure*}[!t]
  \includegraphics[width=\linewidth]{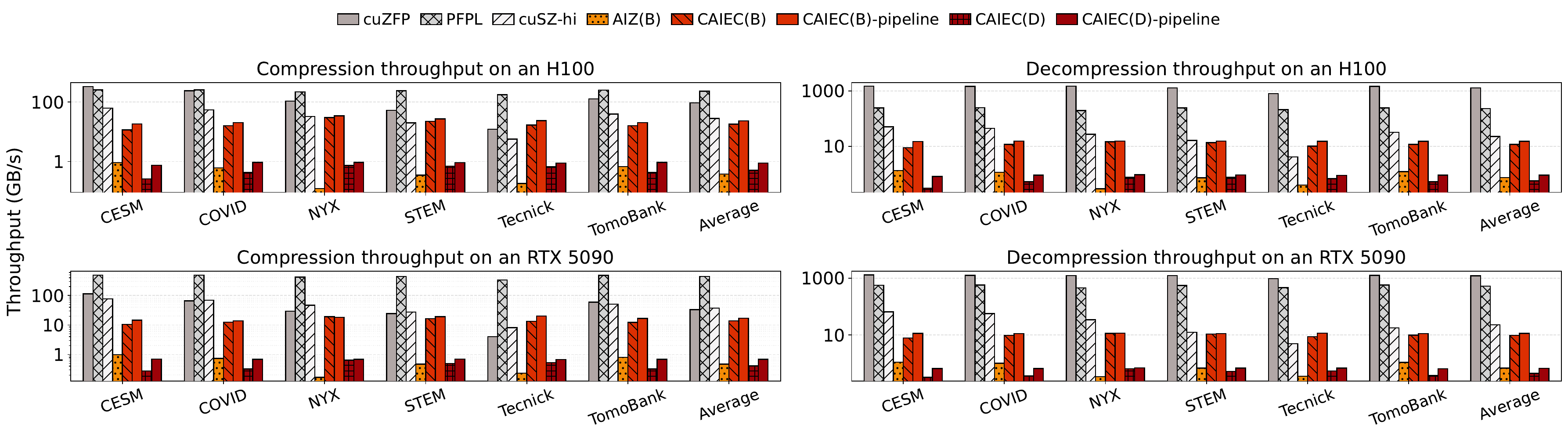}
  \caption{Compression and decompression throughput on an H100 and an RTX 5090 across six datasets.}
  \label{fig:throughput}
\end{figure*}

\noindent\textbf{Two-stream inference--coding pipeline.}
After configuration, CAIEC pipelines inference and coding across blocks. As shown in Fig.~\ref{fig:pipeline}, the TensorRT stream processes block $i+1$ while the ANS stream codes block $i$, with CUDA events enforcing dependencies. Decompression uses the analogous schedule with entropy-decoded latents.

A split GA-ANS API enables asynchronous pipelining: \texttt{create\_session()} prepares persistent buffers, \texttt{launch()} submits encoding kernels, and \texttt{finalize()} synchronizes only when collecting the result. This lets the host launch the next block's inference while coding proceeds.

\subsection{Instantiation on Scientific Data}
\label{sec:adaptation}

CAIEC instantiates its inference--coding workflow on scientific data via lightweight adaptation. These steps make general learned compressors usable for scientific data.


\noindent\textbf{Slice-wise normalization.}
Scientific variables and slices can have substantially different value ranges. CAIEC applies min--max normalization independently to each slice using the convention expected by the target model, and applies the inverse transformation after reconstruction. The corresponding normalization statistics are retained as side metadata.

\noindent\textbf{Model-compatible slice grouping.}
Many learned compression backbones expect a fixed number of input channels. CAIEC groups adjacent scientific slices along the channel dimension to match the target backbone while preserving local cross-slice correlation, without modifying the backbone.

\noindent\textbf{Resolution-aware partitioning and fusion.}
For inputs that exceed memory limits, CAIEC partitions grouped inputs into resolution-aware blocks. It uses overlapped tiling to retain boundary context and reconstructs the output through Hann-window-based weighted fusion~\cite{Pielawski_2020}. The final output is restored to the original layout. Section~\ref{sec:eval} reports the RD and end-to-end impact of these adaptation choices.

%% file: tex/05_evaluation.tex
\section{Experimental Evaluation}
\label{sec:eval}
This section evaluates CAIEC against SOTA compressors in throughput, quality, and scalability, and reports a breakdown quantifying the contribution of each proposed optimization.

\subsection{Experimental Setup}
\textbf{\textit{Platforms.}}
We evaluate CAIEC on two systems: a server with eight NVIDIA H100 (80~GB HBM3) GPUs and two 48-core Intel Xeon Platinum 8468 CPUs, driver 595.71.05, and CUDA Toolkit 12.6; and an NVIDIA RTX~5090 (32~GB) with two Intel Xeon Gold 6530 CPUs, driver 580.82.07, and CUDA 13.0. Both use PyTorch 2.8.0, TensorRT 10.12.0.36, ModelOpt 0.40.0, and ONNX 1.18.0.


\textbf{\textit{Datasets.}}
Table~\ref{tab:datasets} lists five scientific datasets spanning simulations and scientific imaging, plus the Tecnick natural-image dataset. Tecnick provides an in-domain reference for the pretrained image-compression backbones.

\begin{table}[t]
\centering
\caption{Real-world datasets used in the evaluation.}
\label{tab:datasets}
\setlength{\tabcolsep}{3pt}
\renewcommand{\arraystretch}{0.98}
\footnotesize
\resizebox{0.65\linewidth}{!}{
\begin{tabular}{lll}
\hline
\textbf{Dataset} & \textbf{Size} & \textbf{Domain} \\
\hline
CESM~\cite{cesm}     & 1.9 GB & Climate Simulation \\
COVID~\cite{covid2}    & 1.3 GB & Medical X-ray \\
NYX~\cite{sdrb}      & 512 MB       & Cosmology Simulation \\
STEM~\cite{stem}     & 248 MB & Microscope \\
Tecnick~\cite{asuni2014testimages}  & 55 MB        & Photography \\
TomoBank~\cite{DeCarlo_2018} & 640 MB & X-ray Tomography \\
\hline
\end{tabular}
}
\end{table}

\textbf{\textit{Baselines.}}
We instantiate CAIEC on Ball\'e2016~\cite{balle2016end} and DCAE2025~\cite{dcae2025}, denoted CAIEC~(B) and CAIEC~(D), using official pretrained weights. We compare against their unaccelerated implementations and AIZ~\cite{aiz} instantiated on Ball\'e2016 (AIZ~(B)). Rule-based GPU baselines include cuSZ-Hi~\cite{cuszhi}, cuZFP~\cite{cuzfp}, and PFPL~\cite{pfpl}. Latent-coder experiments additionally use nvCOMP~\cite{nvcomp} and DietGPU~\cite{dietgpu}.

\subsection{Throughput}
\textbf{\textit{Overall throughput results.}}
Fig.~\ref{fig:throughput} reports end-to-end compression and decompression throughput. Without pipelining, CAIEC~(B) achieves compression/decompression throughputs of 18.06\slash 11.63~GB/s on the H100 and 13.71\slash 9.70~GB/s on the RTX~5090. On the H100, these results represent 87.01$\times$\slash 77.16$\times$ speedups over the original implementation. CAIEC~(D) achieves 0.503\slash 0.564~GB/s on the H100 and 0.411\slash 0.452~GB/s on the RTX~5090, corresponding to 8.57$\times$\slash 9.94$\times$ speedups on the H100. It remains below cuSZ-Hi's 28.25~GB/s average compression throughput, but substantially narrows the gap between learned and rule-based GPU compressors.

Pipelining overlaps inference on batch $i+1$ with GA-ANS on batch $i$. On the H100, CAIEC~(B/D) reaches 23.40/15.23 and 0.887/0.899~GB/s compression/decompression throughput, with additional speedups of 1.30$\times$/1.31$\times$ and 1.76$\times$/1.59$\times$, respectively. On the RTX~5090, it reaches 16.90/11.43 and 0.687/0.677~GB/s, with speedups of 1.23$\times$/1.18$\times$ and 1.67$\times$/1.50$\times$.

\begin{figure}[t]
  \centering
\includegraphics[width=0.9\linewidth]{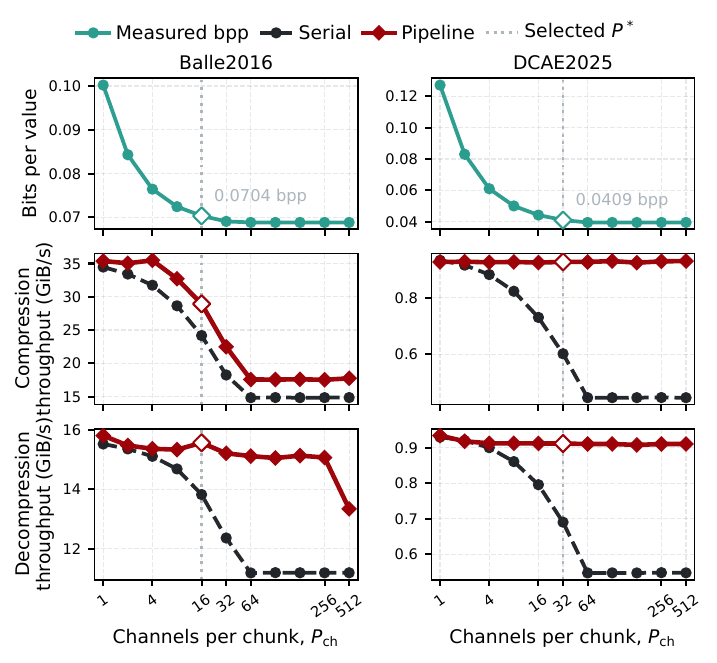}
  \caption{Impact of entropy-coding granularity on compression efficiency and end-to-end throughput for TomoBank.}
  \label{fig:entropy_p_tradeoff}
\end{figure}

\textbf{\textit{Inference-aware entropy-coding granularity.}}
To validate the selection rule in Section~\ref{sec:codesign}, we sweep 11 candidate granularities from $P_{\mathrm{ch}}=1$ to 512 on TomoBank. Fig.~\ref{fig:entropy_p_tradeoff} reports end-to-end bpp together with serial and pipelined compression/decompression throughput. Increasing $P_{\mathrm{ch}}$ reduces bpp until the coding efficiency saturates, but also increases serial coding overhead. For CAIEC~(B), whose inference stage is lightweight, this overhead eventually reaches the pipeline critical path, so pipelined throughput also decreases. For CAIEC~(D), its heavier inference stage hides most of the additional coding cost, keeping pipelined throughput nearly constant despite declining serial throughput.

The selected values balance coding efficiency against end-to-end performance. For CAIEC~(B), $P_{\mathrm{ch}}^\star=16$ reaches 0.0704~bpp while retaining median pipelined compression/decompression throughput of 28.94/15.57~GiB/s; larger values provide little further bpp reduction but noticeably reduce throughput. For CAIEC~(D), $P_{\mathrm{ch}}^\star=32$ reaches 0.0409~bpp and 0.927/0.907~GiB/s pipelined throughput, capturing most of the bpp improvement while the inference-bound pipeline remains stable.

\begin{figure}[H]
  \centering
  \includegraphics[width=\columnwidth]{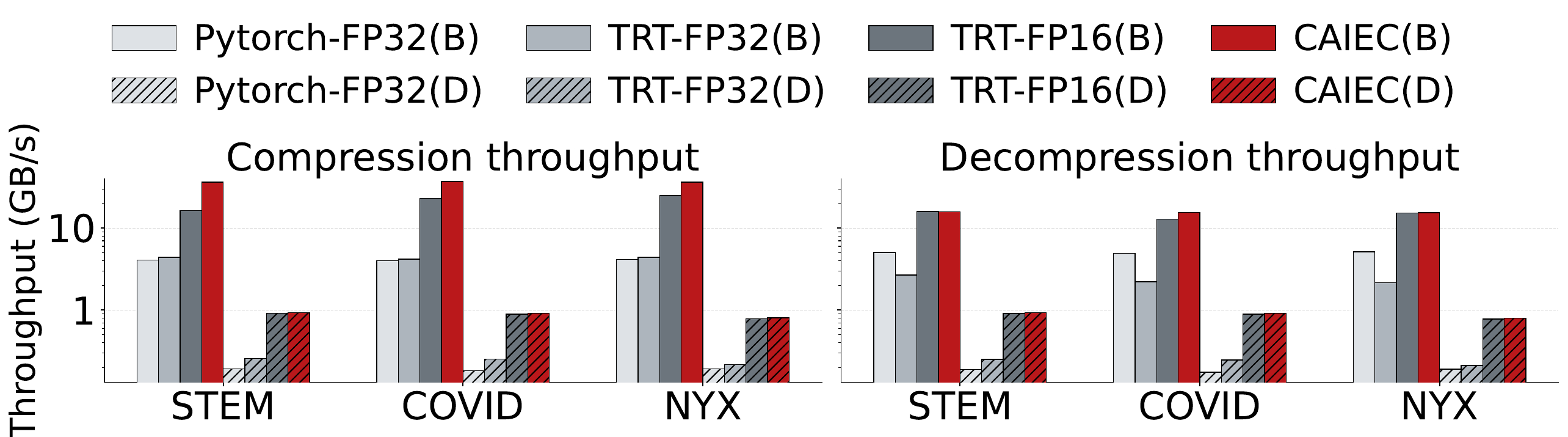}
  \caption{Impact of inference precision on throughput for CAIEC~(B) and CAIEC~(D) across three datasets.}
  \label{fig:precision_thr}
\end{figure}

\textbf{\textit{Impact of mixed precision on throughput.}}
Fig.~\ref{fig:precision_thr} compares PyTorch-FP32, TRT-FP32, TRT-FP16, and CAIEC's mixed precision, separating TensorRT conversion from precision reduction.
Compared with PyTorch-FP32, TRT-FP32 achieves
1.05--1.08$\times$ compression throughput for CAIEC~(B), but only
0.42--0.53$\times$ decompression throughput. TRT-FP16 increases the
corresponding throughputs to 4.03--6.00$\times$ and
2.62--3.16$\times$, while mixed precision reaches
8.77--9.22$\times$ and 3.01--3.15$\times$, respectively. For
CAIEC~(D), TRT-FP32 provides 1.13--1.39$\times$ compression and
1.10--1.40$\times$ decompression throughput. TRT-FP16 achieves
4.07--4.92$\times$ and 4.09--5.08$\times$, whereas mixed precision
reaches 4.19--5.05$\times$ and 4.20--5.20$\times$, respectively. Thus, TensorRT conversion alone provides little benefit; most of the gain comes from reduced-precision execution. 


\begin{figure}[t]
  \centering
  \includegraphics[width=\columnwidth]{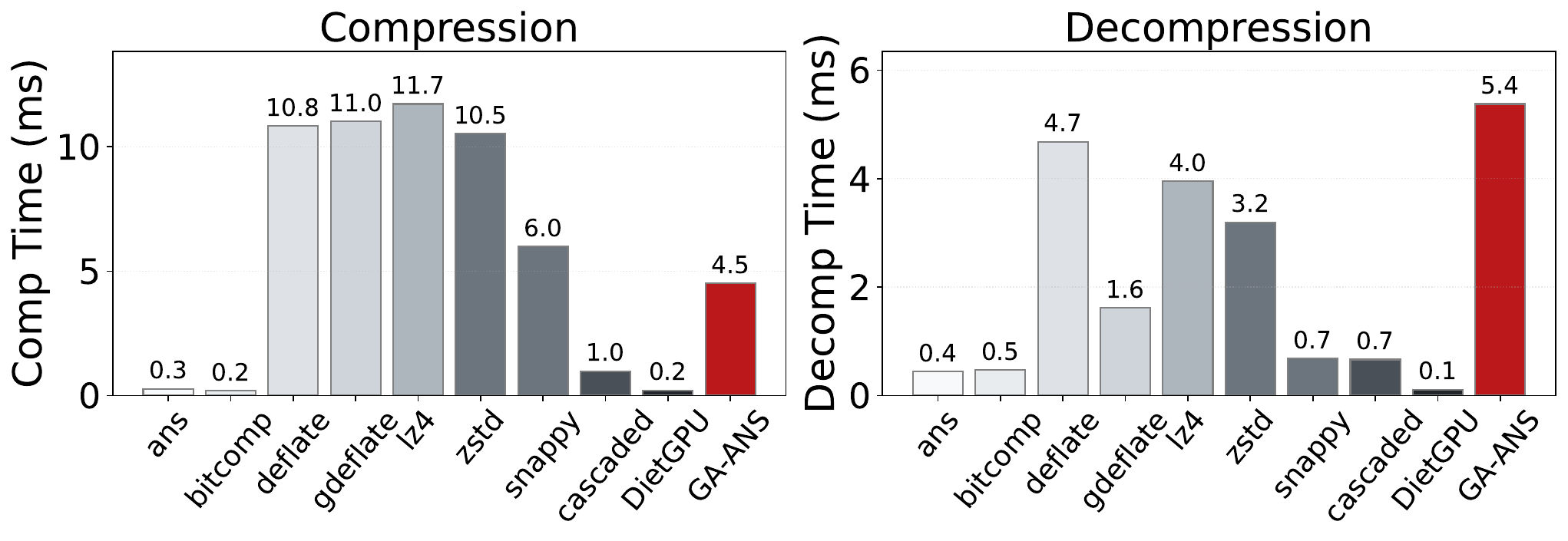}
  \caption{Execution time of GPU-based lossless coders on the latent data of Ball\'e2016 for compression and decompression.}
  \label{fig:latentY_comp_decomp}
\end{figure}

\begin{figure}[t]
  \centering
  \includegraphics[width=.9\columnwidth]{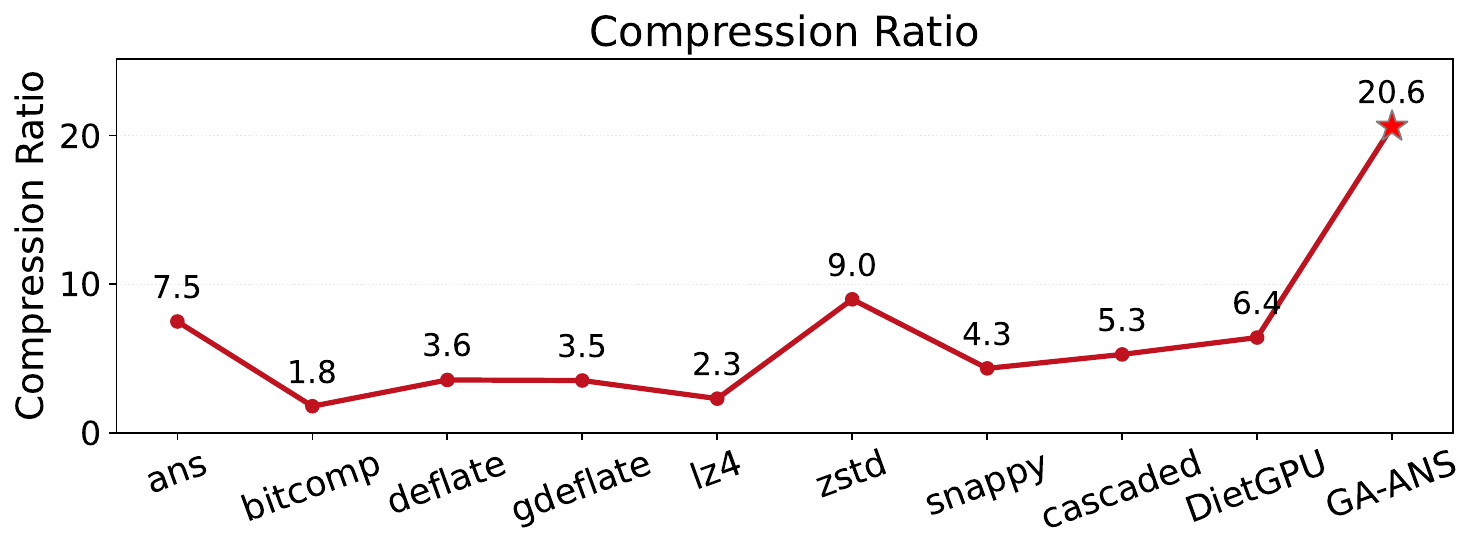}
  \caption{Compression ratio of different GPU-based lossless coders on the latent data produced by Ball\'e2016.}
  \label{fig:latentY_cr}
\end{figure}

\begin{figure}[t]
  \centering
  \includegraphics[width=\linewidth]{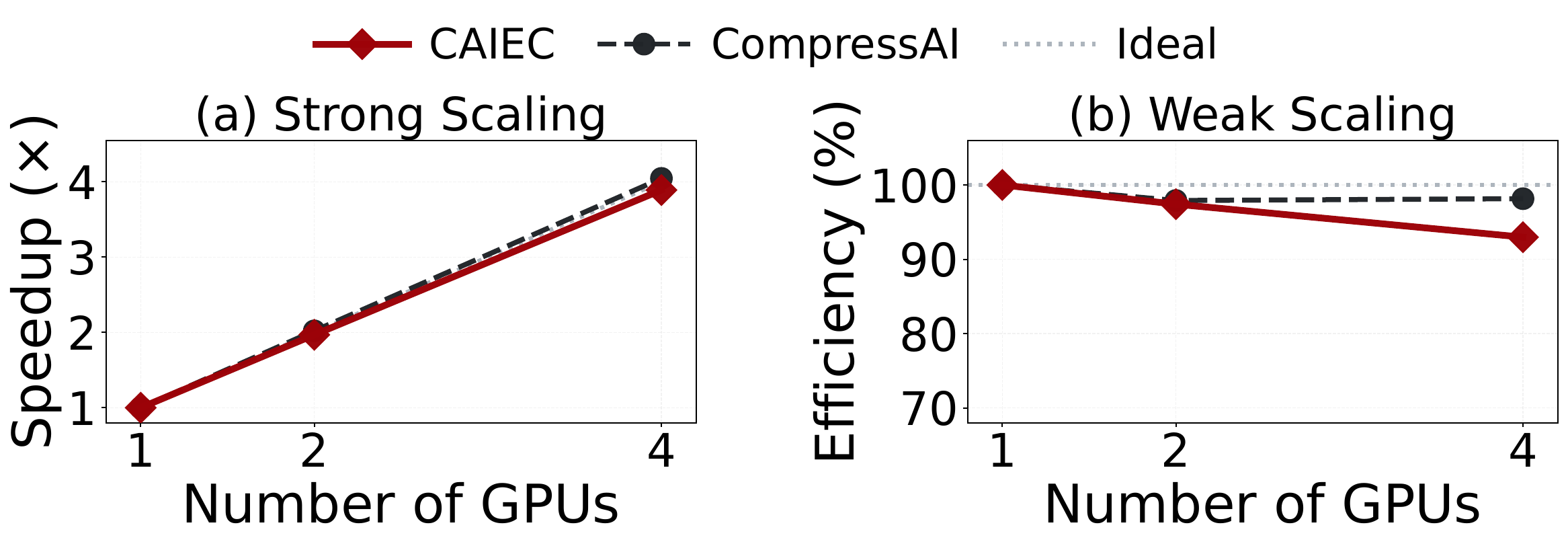}
  \caption{Strong and weak multi-GPU scaling of CAIEC.}
  \label{fig:multi_gpu_scaling}
\end{figure}

\begin{figure*} [!t]
  \includegraphics[width=\linewidth]{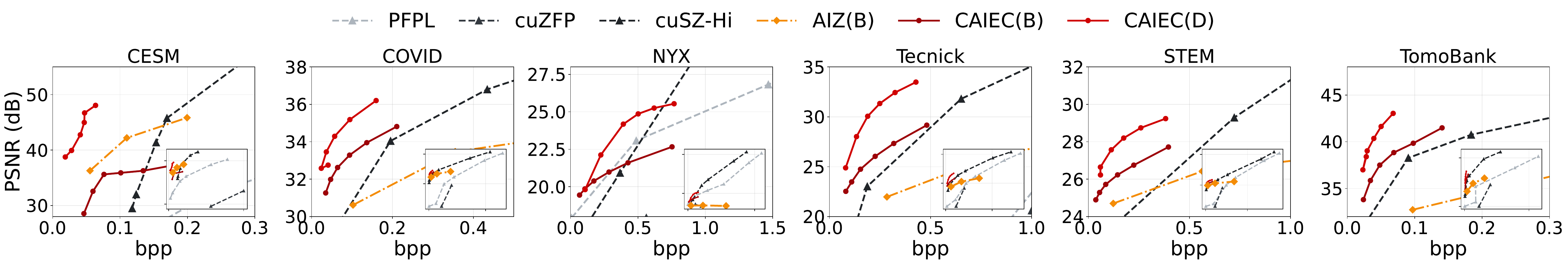}
  \caption{Rate--distortion~(RD) plots showing PSNR versus bitrate across six representative datasets. The main panel in each subfigure zooms into the low-bitrate region, while the inset~(bottom-right) shows the complete RD curves of all methods over the full bitrate range.}
  \label{fig:RD-AI-Non}
\end{figure*}

\textbf{\textit{Throughput--compression trade-off of GA-ANS.}}
Fig.~\ref{fig:latentY_comp_decomp} compares GPU entropy coders on Ball'e2016 latents. GA-ANS achieves 4.52/5.38~ms compression/decompression latency, with compression speed between optimized GPU entropy coders and general-purpose GPU coders.

As shown in Fig.~\ref{fig:latentY_cr}, GA-ANS achieves a 20.6$\times$ compression ratio, substantially higher than existing GPU coders (5.3$\times$--9.0$\times$). Unlike byte-level GPU coders, GA-ANS preserves learned symbol probabilities and reduces partition overhead through coarser-grained coding. Its higher coding latency is therefore an intentional end-to-end trade-off: once coding latency is hidden within the inference budget, preserving latent compression ratio is more valuable than maximizing standalone coder throughput.

\subsection{Multi-GPU Scalability of CAIEC}
Fig.~\ref{fig:multi_gpu_scaling} reports strong and weak scaling for Ball\'e2016 on NYX. Batches are evenly distributed among independent GPU workers without inter-GPU communication. Timings include host-to-device transfer, compression, and decompression, but exclude startup, warmup, and file I/O.

Under strong scaling, the total workload remains fixed at 32 batches. CAIEC reaches 1.97$\times$ speedup on two GPUs and 3.89$\times$ on four GPUs, corresponding to 98.4\% and 97.3\% scaling efficiency. The FP32 baseline reaches 2.02$\times$ and 4.05$\times$, or 101.0\% and 101.1\% efficiency. Overall, both implementations scale nearly linearly because the inputs are independent and require no inter-GPU synchronization.

For weak scaling, each GPU processes eight batches. The baseline achieves 97.9\%/98.2\% efficiency on two/four GPUs, compared with 97.4\%/93.0\% for CAIEC. CAIEC's per-worker coding times remain stable, but host-to-device time increases from 0.230 to 0.261~s, raising round-trip time from 0.434 to 0.467~s. This transfer overhead is more visible for CAIEC because its GPU computation is substantially faster than the baseline's roughly 21~s round trip.

\subsection{Compression Quality}
\textbf{\textit{Overall rate--distortion performance.}}
Fig.~\ref{fig:RD-AI-Non} compares CAIEC with rule-based compressors and AIZ. CAIEC~(D) generally provides the best RD performance because its higher-capacity transforms capture complex data structures more effectively. The lighter CAIEC~(B) still outperforms rule-based methods over a broad bitrate range, showing that its throughput gains do not eliminate the learned model's RD advantage. CAIEC~(B) also requires fewer bits than AIZ~(B) at the same PSNR, particularly at low bitrates, because GA-ANS retains more of the entropy model's latent-coding efficiency.

\begin{figure}[t]
  \centering
  \includegraphics[width=\columnwidth]{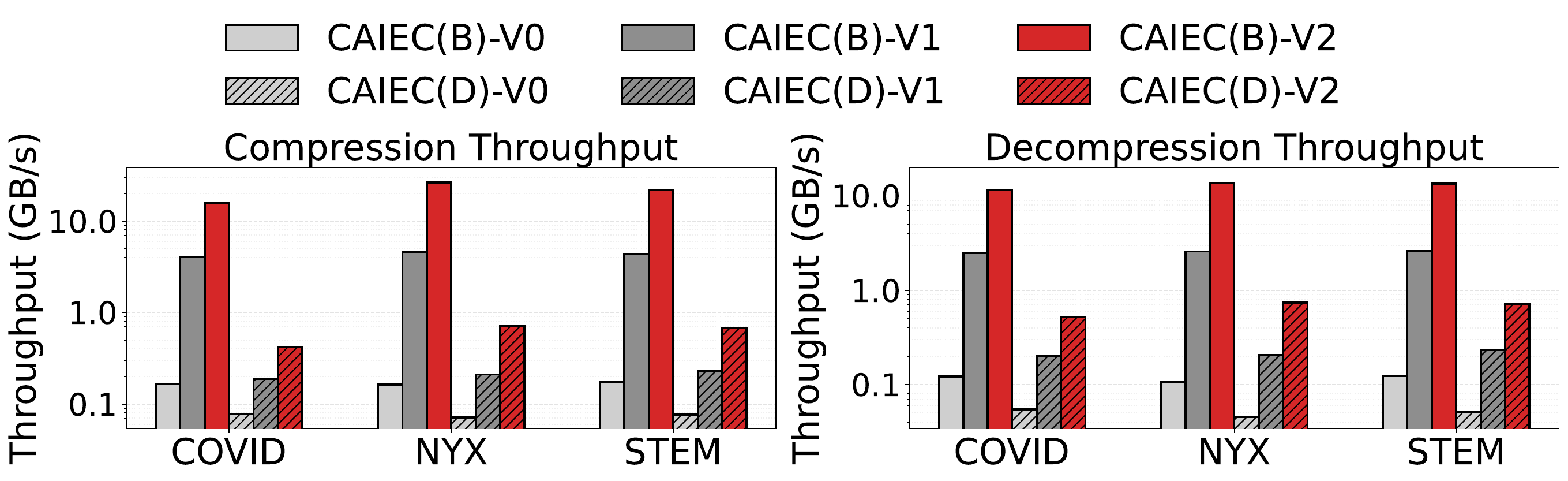}
  \caption{Compression and decompression throughput of CAIEC~(B) and CAIEC~(D) under progressive optimizations.}
  \label{fig:ablation0_thr}
\end{figure}

\begin{figure}[t]
  \centering
  \includegraphics[width=\columnwidth]{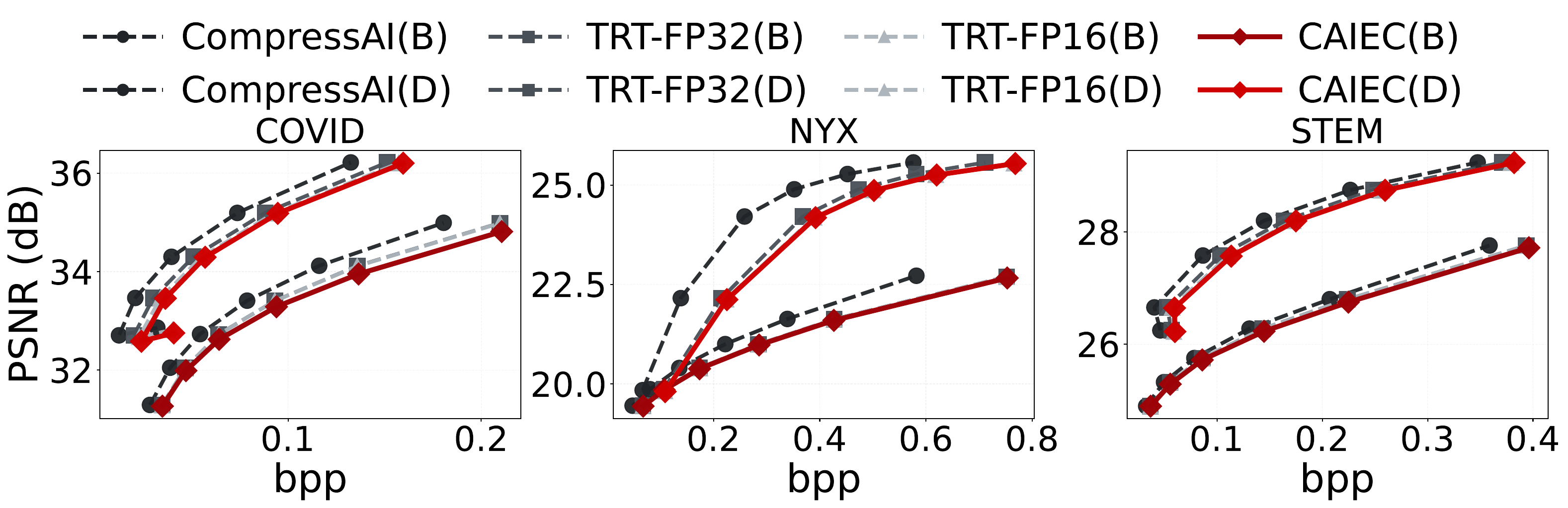}
  \caption{Rate--distortion comparison under different precision settings for CAIEC~(B) and CAIEC~(D) on three datasets.}
  \label{fig:precision_rd}
\end{figure}

\textbf{\textit{Rate--distortion impact of mixed-precision inference.}}
\label{caiecwithFP32FP16MIX}
Fig.~\ref{fig:precision_rd} compares four configurations for both backbones. CompressAI is the original FP32 implementation with CPU ANS, whereas TRT-FP32 and TRT-FP16 run the same pretrained models in TensorRT at the indicated uniform precision and replace CPU ANS with GA-ANS.

Across the three representative datasets, TRT-FP32 with GA-ANS closely follows the original CompressAI curves: PSNR is unchanged, and only small bpp differences are visible. This indicates that GA-ANS retains the learned entropy model's coding effectiveness, with only limited bitrate impact from its chunked bitstream organization. CAIEC also remain close to the FP32 references. In particular, CAIEC introduces only small PSNR changes relative to the uniform-precision baselines despite assigning FP8 to selected components, showing that component-wise sensitivity analysis avoids aggressively quantizing numerically sensitive modules. Together, these results show that GA-ANS preserves latent coding efficiency while CAIEC's mixed-precision inference controls reconstruction degradation, thereby maintaining the pretrained models' overall RD performance.
\begin{figure}[t]
  \centering
  \includegraphics[width=\columnwidth]{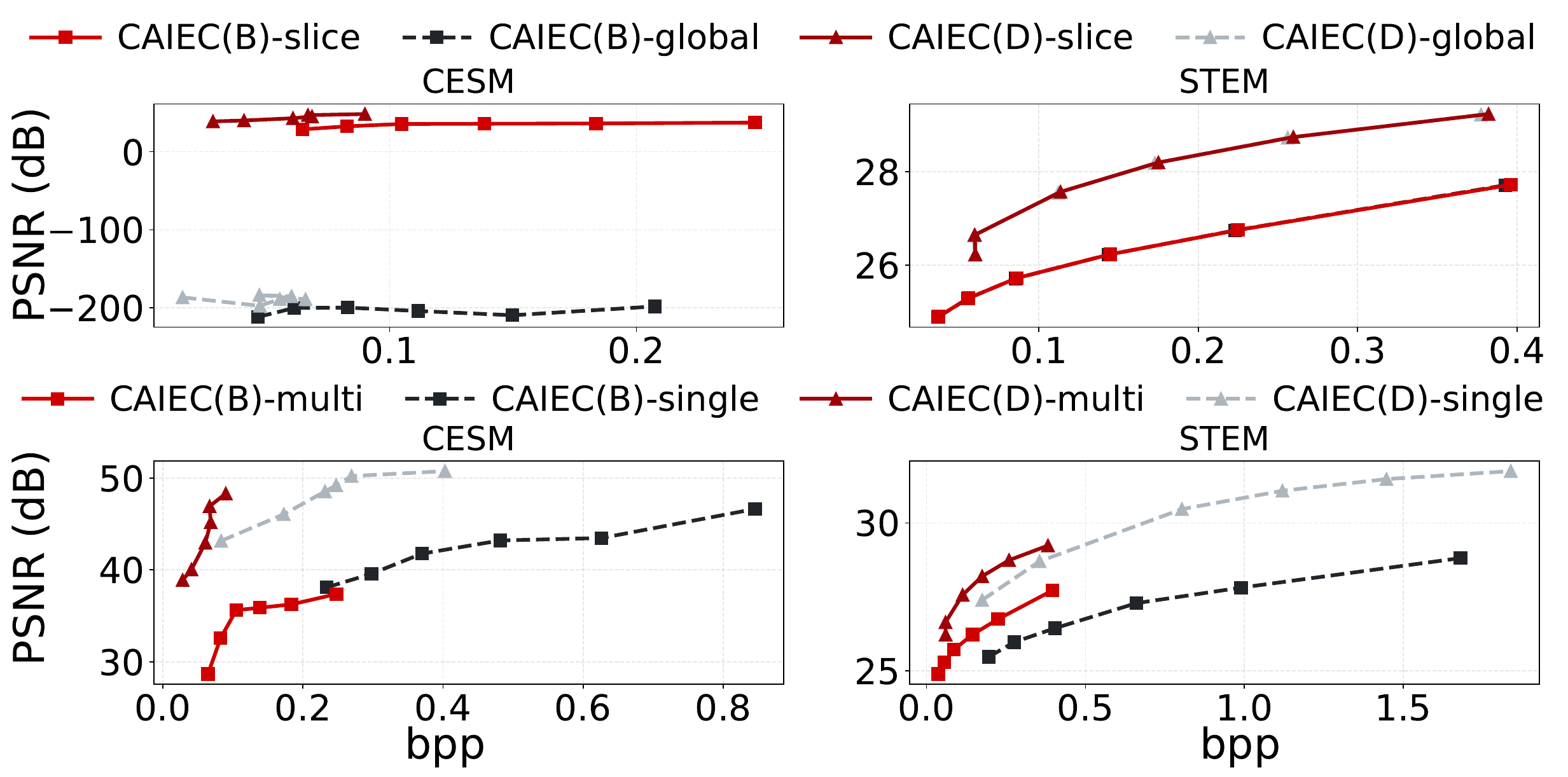}
  \caption{RD curves for the ablation of slice-wise normalization and multi-channel stacking on CESM and STEM.}
  \label{fig:ablationAB_rd}
\end{figure}

\begin{figure}[t]
  \centering
  \includegraphics[width=\columnwidth]{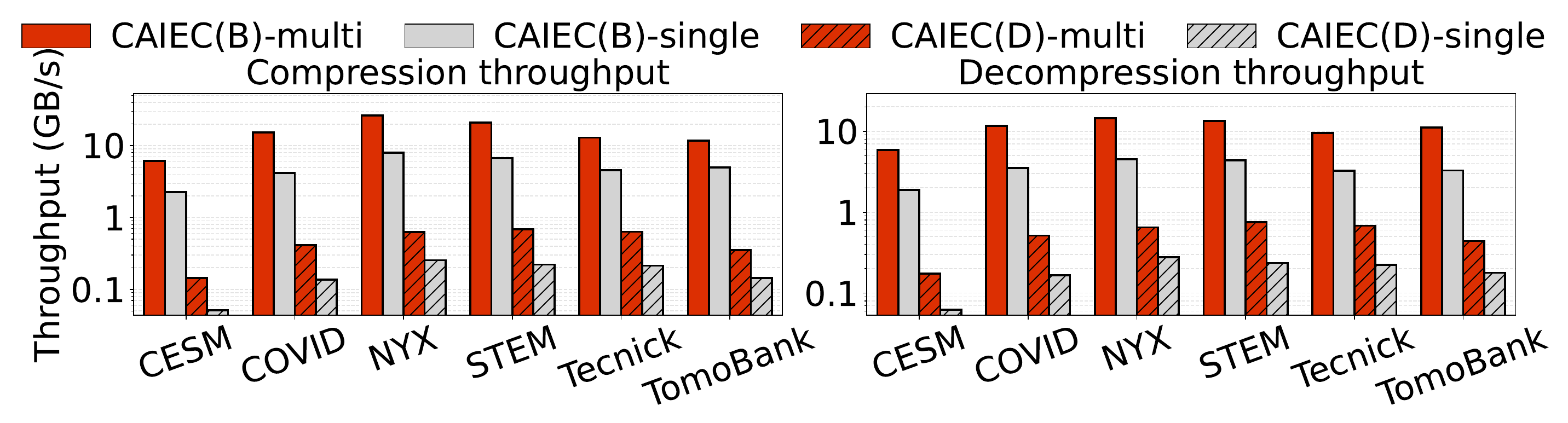}
  \caption{Throughput of multi-channel and single-channel for CAIEC~(B) and CAIEC~(D) across multiple datasets.}
  \label{fig:ablationB_thr}
\end{figure}

\subsection{Ablation Studies}
\textbf{\textit{Effectiveness of end-to-end optimizations.}}
Fig.~\ref{fig:ablation0_thr} compares full precision with CPU entropy coding (V0), GA-ANS acceleration (V1), and mixed precision on top of GA-ANS (V2). V1 improves compression/decompression throughput by 24.49--28.04$\times$/20.38--24.26$\times$ for CAIEC~(B) and 2.42--2.96$\times$/3.73--4.53$\times$ for CAIEC~(D). V2 further provides 3.91--5.78$\times$/4.69--5.33$\times$ and 2.22--3.42$\times$/2.54--3.60$\times$ gains over V1. This result motivates balancing the two stages instead of applying a fixed acceleration strategy across backbones.

\textbf{\textit{Effectiveness of slice-wise normalization.}}
Fig.~\ref{fig:ablationAB_rd} compares slice-wise and global normalization on CESM and STEM. Slice-wise normalization improves RD on both datasets and avoids the negative PSNR of global normalization on CESM. CESM slices have widely different value ranges, so one global range is dominated by large-magnitude slices and under-resolves the others. Normalizing each slice independently removes this imbalance before compression.

\textbf{\textit{Effectiveness of multi-channel stacking.}}
Multi-channel stacking improves PSNR at the same bitrate on CESM and STEM (Fig.~\ref{fig:ablationAB_rd}) and increases compression/decompression throughput by 2.36--3.65$\times$/2.91--3.37$\times$ for CAIEC~(B) and 2.45--3.10$\times$/2.34--3.19$\times$ for CAIEC~(D)(Fig.~\ref{fig:ablationB_thr}). Stacking exposes local inter-slice correlation to the transform and fills the backbone's multi-channel input in one inference call; processing slices independently loses this redundancy and repeats launch and model-execution overhead.

\begin{figure}[t]
  \centering
  \includegraphics[width=\columnwidth]{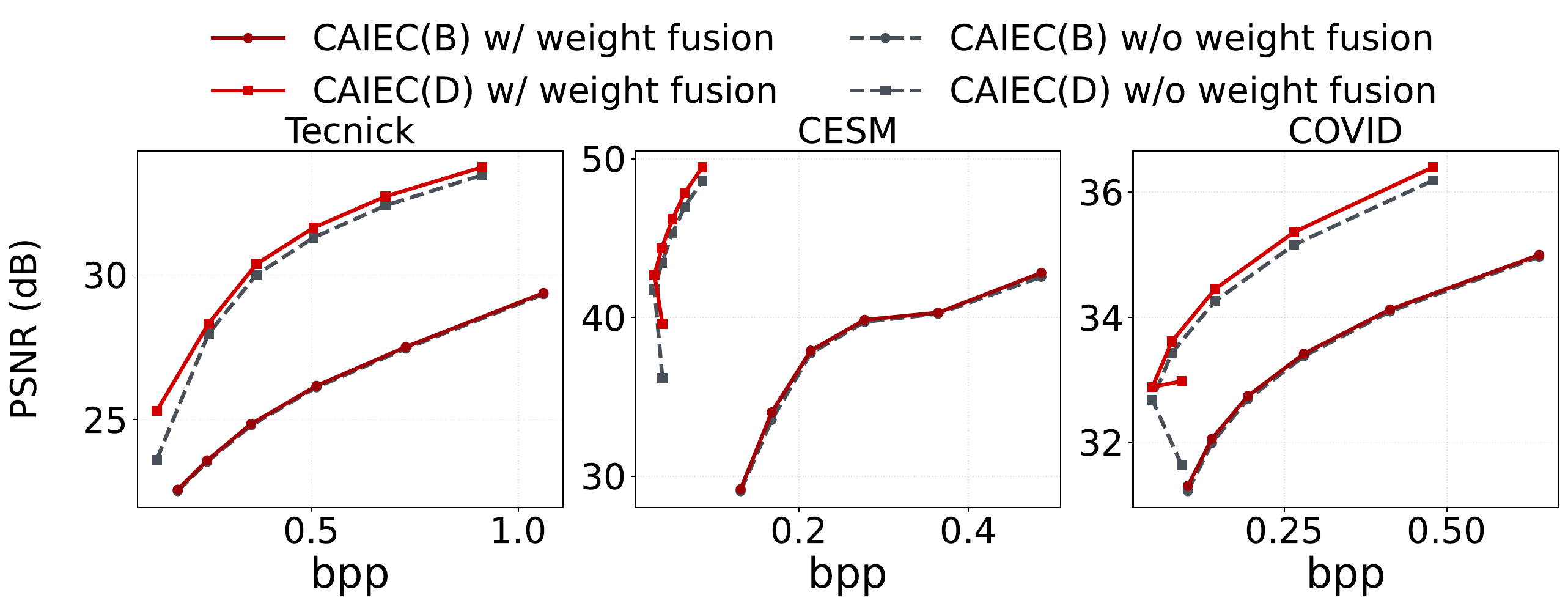}
  \caption{RD curve comparison of CAIEC~(B) and CAIEC~(D) with and without weighted fusion on Tecnick, CESM, and COVID.}
  \label{fig:ablationC}
\end{figure}

\textbf{\textit{Effectiveness of weighted fusion.}}
Fig.~\ref{fig:ablationC} shows that weighted fusion improves PSNR at comparable bitrates by about 1~dB on Tecnick and COVID and up to 3~dB on CESM. Direct stitching gives boundary pixels the same weight as block interiors even though they have less spatial context and larger reconstruction error. Hann-window weighting suppresses these unreliable boundaries and favors overlapping block centers, explaining the larger gain on CESM.

%% file: tex/06_application.tex
\section{Use Case Study of CAIEC}
\label{sec:applications}

We further apply CAIEC to CRA5~\cite{cra5} to evaluate its generalization beyond generic learned compression backbones. CRA5 is a domain-specific climate compressor trained on ERA5~\cite{hersbach2020era5} reanalysis data, and compresses over 400\, TiB of hourly ERA5 fields (1973--2023) to roughly 1\, TiB. To assess downstream usability, we drive MPTRAC~\cite{hoffmann2021mptrac} particle trajectory simulations with reconstructed ERA5 data and compare CAIEC with the rule-based lossy compressor cuSZ-Hi at matched throughput and compression ratio.

Fig.~\ref{fig:cra5_analysis} shows that original CRA5 achieves the highest downstream accuracy over a continuous 8-day period, but only achieves 2.1/1.4~GB/s compression/decompression throughput. With CAIEC, CRA5 reaches 20.4/13.7~GB/s at a similar CR (446$\times$ vs. 445$\times$), with only minor accuracy degradation. Although cuSZ-Hi achieves higher throughput (53.7/41.1~GB/s) at a similar CR of 439$\times$, it yields lower downstream accuracy. Thus, CAIEC makes high-fidelity learning-based compression substantially more practical by delivering nearly an order-of-magnitude throughput improvement while preserving most of the original model’s scientific accuracy.

%% file: tex/07_discuss.tex
\section{Discussion}
\label{sec:discuss}
  
\textbf{Parallelism hierarchy and bottleneck shifting.}
CAIEC exploits parallelism across scientific-data batches, inference and coding stages, GA-ANS chunks, warp lanes, and GPUs. These forms are complementary rather than interchangeable. Finer coding granularity increases inter-chunk parallelism but adds bitstream overhead, whereas coarser granularity preserves compression ratio and relies on warp-level parallelism. Accelerating one level can expose another bottleneck: mixed-precision inference can make entropy coding dominant, while fast multi-GPU execution can expose host data movement. End-to-end acceleration therefore requires pipeline-wide coordination of parallelism rather than independent stage optimization.

\textbf{Configuration overhead and amortization.}
CAIEC shifts part of the optimization cost to offline configuration. On CESM, exhaustively evaluating 2,048 precision configurations takes approximately two days, whereas component-wise profiling and guided recovery complete in a few minutes. This cost is amortized when the same backbone and platform repeatedly process data with similar characteristics. However, changes in the target GPU, quality budget, or workload distribution may require re-profiling. CAIEC therefore best suits repeated or large-scale deployments, while its configuration overhead may be less attractive for one-shot compression of small inputs.

\begin{figure}[t]  
  \centering
    \includegraphics[width=\linewidth]{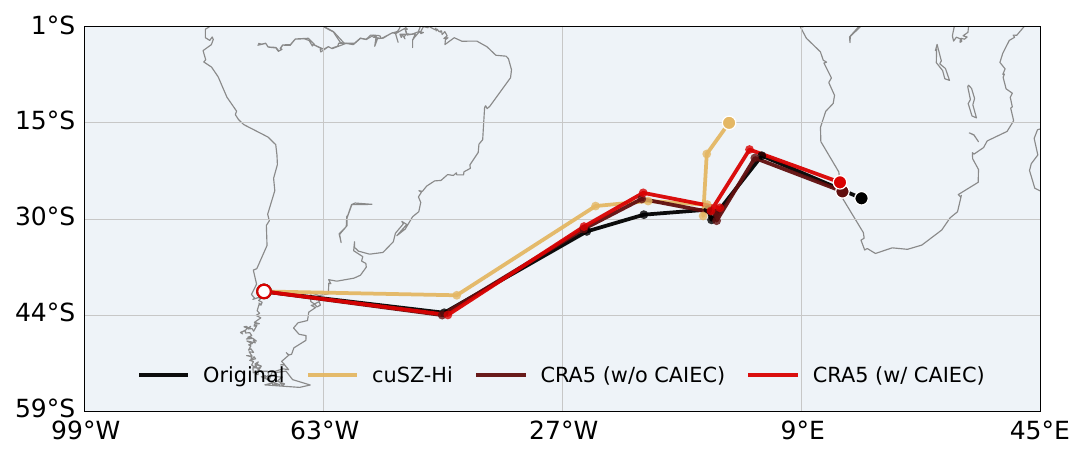}
    \caption{MPTRAC-based downstream analysis.}
    \label{fig:cra5_analysis}
  \end{figure}

\textbf{Implications for compression model design.}
CAIEC removes substantial inference, coding, and stage-serialization overhead but cannot eliminate the intrinsic cost of a complex backbone. Even with pipelining, DCAE2025 reaches only 0.887/0.899~GB/s compression/decompression throughput on the H100, leaving a substantial gap from lightweight learned and rule-based compressors. Future learned compressors should therefore pursue \emph{model--inference--coding co-design}: beyond RD performance, their architectures should favor GPU-efficient operators, balanced inference stages, regular latent layouts, and representations supporting parallel entropy coding with limited partition overhead. Domain structure and physical priors may further reduce model complexity, but their system-level benefits require joint evaluation with inference and coding.

%% file: tex/08_conclusion.tex
\section{Conclusion and Future Work}
\label{sec:conclusion}

In this work, we proposed CAIEC, an end-to-end acceleration framework for learning-based scientific compressors. It jointly optimizes mixed-precision inference and entropy coding, enabling controllable throughput--quality trade-offs. CAIEC reaches average compression and decompression throughputs of 23.40\,GB/s and 15.23\,GB/s, up to 112.72$\times$ and 101.04$\times$ faster than non-accelerated baselines, and attains higher downstream accuracy than cuSZ-Hi at comparable compression ratios on a climate HPC application. CAIEC thus makes learned scientific compression practical at HPC scale, without trading quality for throughput.

Two directions remain for future work: extending CAIEC to other domains where storage pressure and fidelity requirements favor learned compression, and integrating it into storage system I/O paths so that learned compression stays transparent to applications.

%% file: tex/09_ack.tex
\section{Acknowledgement}
This work was supported by the Innovation Funding of ICT, CAS (Grant No. E461050), the National Key Research and Development Program of China (Grants No. 2020YFA0907000 and 2025YFB30037002), the National Natural Science Foundation of China (Grant Nos. 62032023 and T2125013), and Ant Group. The work was also supported by International Partnership Program of Chinese Academy of Sciences (Grant No. 094GJHZ2024019MI). The AI-driven experiments, simulations and model training were performed on the robotic AI-Scientist platform of Chinese Academy of Sciences.